\documentclass{article}

\usepackage{PRIMEarxiv}

\usepackage[utf8]{inputenc} 
\usepackage[T1]{fontenc}    
\usepackage{hyperref}       
\usepackage{url}            
\usepackage{booktabs}       
\usepackage{amsfonts}       
\usepackage{nicefrac}       
\usepackage{microtype}      
\usepackage{lipsum}
\usepackage{fancyhdr}       
\usepackage{graphicx}       
\usepackage{amsmath}
\graphicspath{{media/}}     
\usepackage{algorithm}
\usepackage{algpseudocode}
\usepackage{amsmath}

\title{Latent Anomaly Detection: Masked VQ-GAN for Unsupervised Segmentation in Medical CBCT}

\author{
Pengwei Wang \textsuperscript{1}\\
\textsuperscript{1}Department of Mechanical Engineering \\ College of Design and Engineering \\ National University of Singapore\\
}

\begin{document}
\maketitle

\begin{abstract}
Advances in treatment technology now allow for the use of customizable 3D-printed hydrogel wound dressings for patients with osteoradionecrosis (ORN) of the jaw (ONJ). Meanwhile, deep learning has enabled precise segmentation of 3D medical images using tools like nnUNet.
However, the scarcity of labeled data in ONJ imaging makes supervised training impractical. This study aims to develop an unsupervised training approach for automatically identifying anomalies in imaging scans.
We propose a novel two-stage training pipeline. In the first stage, a VQ-GAN is trained to accurately reconstruct normal subjects. In the second stage, random cube masking and ONJ-specific masking are applied to train a new encoder capable of recovering the data.
The proposed method achieves successful segmentation on both simulated and real patient data.
This approach provides a fast initial segmentation solution, reducing the burden of manual labeling. Additionally, it has the potential to be directly used for 3D printing when combined with hand-tuned post-processing.
\end{abstract}

\section{Introduction}
\label{sec:introduction}
Osteoradionecrosis of the jaw (ORN) is a debilitating complication that arises following radiation therapy for head and neck malignancies \cite{marx1983osteoradionecrosis}. It involves the necrosis of previously irradiated bone, culminating in chronic, non-healing wounds that substantially increase the risk of infection and other significant morbidities \cite{regaud1922sensibilite}. Historically, the management of ORN has relied on invasive surgical interventions, which impose considerable physical and psychological burdens on patients \cite{annane2004hyperbaric}. Recent advancements in medical technology, particularly the utilization of customizable 3D-printed hydrogel wound dressings, offer promising, minimally invasive alternatives that may enhance therapeutic outcomes while reducing patient discomfort \cite{vijayavenkataraman20183d}.

In parallel, developments in deep learning have provided powerful tools for medical image analysis, particularly for segmentation tasks. One such tool, nnUNet \cite{isensee2021nnu}, has demonstrated success in segmenting 3D medical images with high accuracy. However, the field of ORN imaging faces a major challenge due to the scarcity of labeled datasets, making supervised learning approaches difficult to implement. To address this issue, this study proposes an unsupervised approach aimed at identifying anomalies in patient imaging scans. The development of such methods is crucial for reducing the dependency on manual labeling and accelerating the treatment planning process.

This project employs unsupervised anomaly detection to segment anomalous regions in cone-beam computed tomography (CBCT) scans of patients with osteonecrosis of the jaw (ONJ). Specifically, a Vector Quantized Generative Adversarial Network (VQ-GAN) \cite{esser2021taming} is trained using a novel methodology, achieving high-fidelity reconstruction of the original anatomical structures, including the likely pre-morbid dentition. A subsequent post-processing pipeline is implemented to generate 3D-printable models of the affected regions. The key contributions of this work are summarized as follows:

\begin{enumerate}
    \item To the best of my knowledge, this is the first study to apply unsupervised anomaly detection to CBCT dental imaging data.

    \item A novel training protocol is introduced to address the challenges associated with reconstruction errors and the failure to recover extensive anatomical anomalies in dental imaging data. This protocol is validated on both simulated anomalies and one actual patient scan.

    \item A comprehensive post-processing pipeline is developed to produce 3D-printable models of the wound area, offering potential utility in personalized surgical planning and intervention.
\end{enumerate}

\section{Related Works}
\label{sec:related}
\subsection{Representation Learning}
Representation learning aims to automatically extract useful features from raw data to support efficient model learning. Traditional manual feature extraction often fails with complex data like images and audio, while neural networks effectively learn hierarchical, abstract representations. Architectures such as AutoEncoders (AEs) \cite{rumelhart1986learning}, Variational AutoEncoders (VAEs) \cite{kingma2013auto}, and Vector Quantized VAEs (VQ-VAEs) \cite{van2017neural} are essential: AEs perform dimensionality reduction, VAEs add probabilistic modeling for generative tasks, and VQ-VAEs use discrete latent spaces for sharper reconstructions. These methods are highly relevant for Unsupervised Anomaly Detection (UAD), most of the UAD methods leverage representation learning models as their base model.

\subsubsection{AutoEncoder}
AutoEncoders (AE) are a kind of neural network to learn a compressed representation for a given data distribution \cite{rumelhart1986learning}. A encoder \(f_e\) is used to compress data \(x\) into latent representation \(z = f_e(x)\) and a decoder is used to recover the data \(\hat{x} = f_d(z)\). To train the AE, the basic loss function cam be a simple mean squared error \(\mathcal{L}_{MSE} = \|x - \hat{x}\|_2^2\). This basic setting is useful, but have a set of problems, including blurry recovered image due to the MSE function, and a unstructured latent space, where there is no method to direct sample in its latent space and generate new sample by the decoder.

\subsubsection{Variational AutoEncoder}
Variational AutoEncoder (VAE) \cite{kingma2013auto}, impose a probabilistic structure on the latent space. The encoder in VAE map the input data into a distribution over the latent space, parameterized by mean \(\mu\) and variance \(\sigma^2\), from which a latent representation \(z\) in sampled. Then the decoder can recover the data using the sampled latent vector. The loss term for VAE, in addition to the MSE loss, has another regularization term to ensure the distribution of the latent space is close to a multivariate Gaussian distribution, usually Kullback-Leibler (KL) divergence chosen. Essentially, the encoder network parameterises a posterior distribution \(q(z|x)\) over the latent variables conditioned on the input data, the loss of KL divergence ensures that this posterior approximates the prior distribution \(p(z)\), and the decoder models the conditional distribution \(p(x|z)\). By enforcing a Gaussian structure in the latent space, VAEs facilitate the interpolation and generation of new data samples that are coherent and varied, enhancing the model's utility in generative tasks. The incorporation of the KL divergence term provides a regularizing effect, making VAEs generally more robust to overfitting compared to traditional autoencoders. However, The Gaussian assumption and the MSE loss often lead to blurring in the reconstructed outputs. This is particularly evident in applications like image reconstruction where sharpness and detail are crucial. In some cases, VAEs can experience mode collapse, where the model ignores certain modes of the data distribution, leading to less diversity in the generated samples.

\subsubsection{The Vector Quantized Variational AutoEncoder}
The Vector Quantized Variational AutoEncoder (VQ-VAE), following the variational autoencoder idea, uses a discrete latent representation \cite{van2017neural}. In the VQ-VAE, after getting a code from the encoder, it is seen as a set of feature vectors, which are going to be replaced by their nearest neighbour in a predefined codebook, a fixed collection of feature vectors. This quantization step replaces the Gaussian sampling process seen in standard VAEs, thereby making the latent space discrete. The decoder uses this quantized vector to reconstruct the input data.

The loss function for VQ-VAE is still similar to VAE, including a reconstruction loss term, typically MSE. For the regularization term for the latent space, the loss encourage the encoder's output to be close to the selected codebook vector, and also encourage the codebook vector be closer to the encoded feature, which can be written as \cite{van2017neural}:
\begin{align}
    \mathcal{L}_{VQ} = \|sg[f_e(x)]-e\|_2^2 + \beta \|f_e(x)-sg[e]\|_2^2
\end{align}
where \(sg[\cdot]\) means stop gradient. These two terms essentially bring the codebook and encoded vector close together. The prior distribution of the latent space is assumed to be uniform, and the code is directly queried by the nearest vector not sampled, the KL divergence is constant w.r.t. the encoder parameters. As a result, the KL divergence term is ignored in VQ-VAE.

By using discrete codebook vectors, VQ-VAE often yields sharper reconstructions than traditional VAE. Furthermore, VQ-VAEs can efficiently encode information, as each vector in the codebook can represent a large and complex pattern within the data, making them particularly useful in tasks like speech and image synthesis where high fidelity is crucial.

\subsection{Unsupervised Anomaly Detection}
For pixel-level anomaly detection, i.e. anomaly segmentation, there are mainly two categories of methods \cite{yang2021visual, lagogiannis2023unsupervised}: image reconstruction and feature modeling.

\subsubsection{Image reconstruction}
Image reconstruction-based UAD focuses on reconstructing normal images and detecting anomalies based on the reconstruction error. Formally, these methods learn from a healthy distribution \(x \in \mathbb{R}^{D \times H \times W}\) and optimize \(argmin \|M(x) - x\|\). The main difference in these methods are the image reconstruction techniques, where a large variety of generative model can be used. For example, VAE \cite{kingma2013auto} is a classical choice which can be further scaled up while keeping the same overall architecture. GAN \cite{goodfellow2014generative} is used in the model f-AnoGAN \cite{schlegl2019f}. A transformer-based autoencoder is proposed by Ghorbel et al \cite{ghorbel2022transformer} with strong performance. Ideally, with a perfect reconstruction, the input anomaly can be easily identified, while in reality, this method is largely affected by reconstruction error, leaving the anomaly segmentation noisy.

\subsubsection{Feature modeling}
Feature modeling methods mainly focus on detecting the anomaly in a feature space, which is often defined by a pretrained network such as ResNet \cite{he2016deep}. Some recent methods are like Reverse Distillation (RD) \cite{deng2022anomaly} and Feature AutoEncoder (FAE) \cite{meissen2022unsupervised}, they all try to reconstruct the multi-layer feature map extracted by a pretrained network, and generate the anomaly map by resizing the anomaly map in different scale back to the image resolution. Specifically, RD leverage a teacher-student structure, the training objective is to minimize the distance of the feature maps in each layer of the student and teacher. However, different from the normal distillation structure, where student also have the same input as the teacher, the student in RD serves as a decoder, decoding the information passed by the teacher. Such bottleneck design help the student to capture the most essential information, and finally help the anomaly detection. FAE have a more straightforward structure, the overall structure of FAE is still a autoencoder, only the input is replaced by a multi-layered feature map extracted by the network and resized into the same size. The training objective is to minimize the SSIM \cite{wang2004image} between the input and output. Feature modeling based method alleviate the problem of pixel level reconstruction error and currently perform better than the image reconstruction method, as is reported by Lagogiannis et al. \cite{lagogiannis2023unsupervised}. However, on one hand, the reconstruction error actually still exists, only in the feature space.

\section{Methods}
\label{sec:method}
\begin{figure}
    \centering
    \includegraphics[width=1\linewidth]{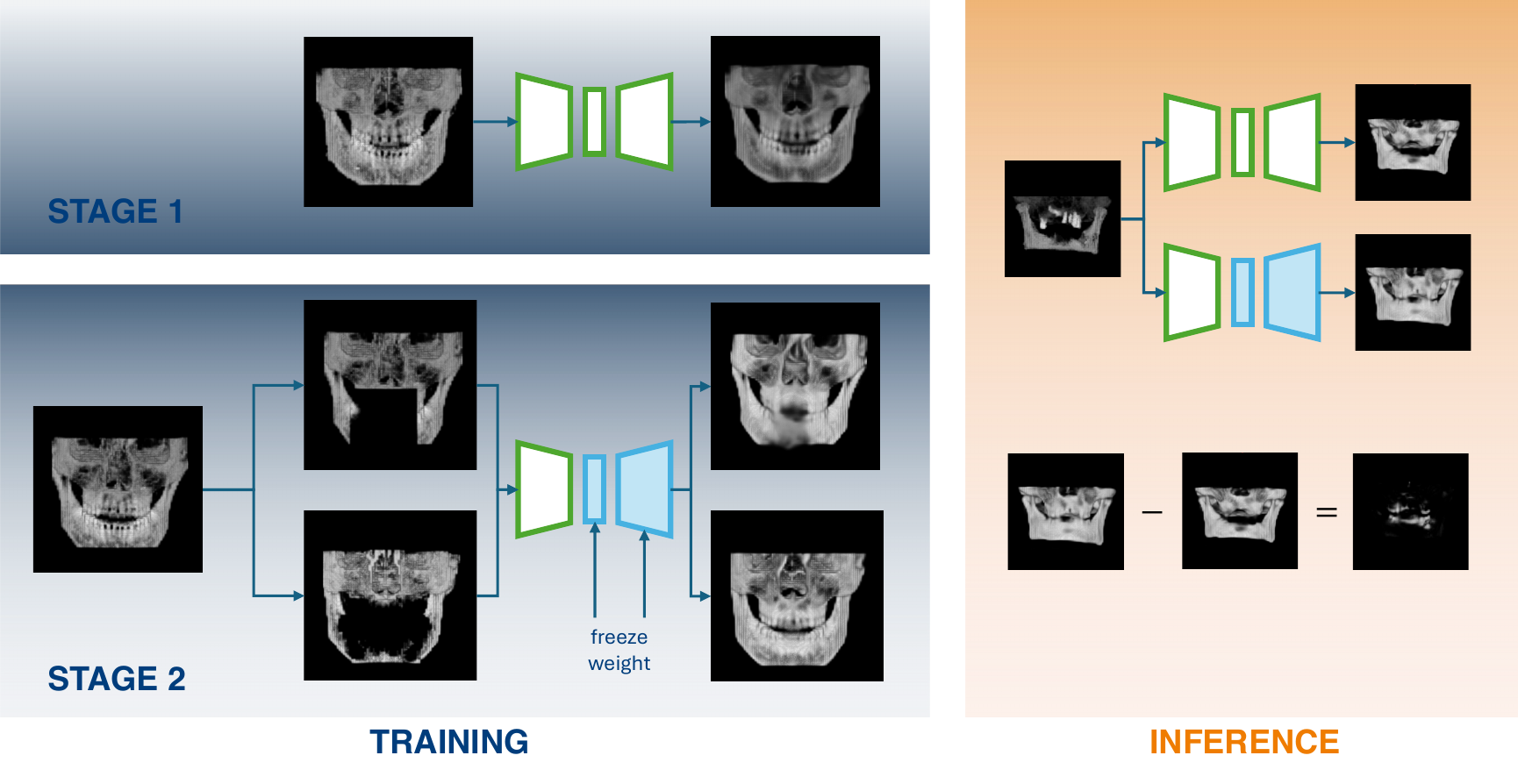}
    \caption{Overview of the proposal anomaly detection scheme}
    \label{fig:overview}
\end{figure}

\subsection{Training Scheme}
The overall method is shown in \ref{fig:overview}. The training is divided into two stages, all trained on a healthy dataset \(x \in \mathbb{R}^{D \times H \times W}\). In the first stage, the model is trained to reconstruct a faithful result from the image by learning to project and recover from a lower dimension \(z \in \mathbb{R}^{a \times b \times c} \), the training objective can be varied depending on the specific model used. In the second stage, the codebook and decoder weights are frozen and only the encoder is trained, meaning a fixed latent space. Two masking methods are used, the first one is a random cube with different sizes, and the second one focuses on the teeth area and masking all the surrounding region so that the model learns to recover from an image with incomplete information.

During the inference, both models trained in two stages are used to generate two reconstructions. The first reconstructed image will fail on the part with ONJ, and the second reconstruction will try to recover an image with all teeth and jaw generated. By taking the difference between these two, the anomaly is clearly segmented.

\subsection{Network: VQ-GAN}
\begin{figure}
    \centering
    \includegraphics[width=0.85\linewidth]{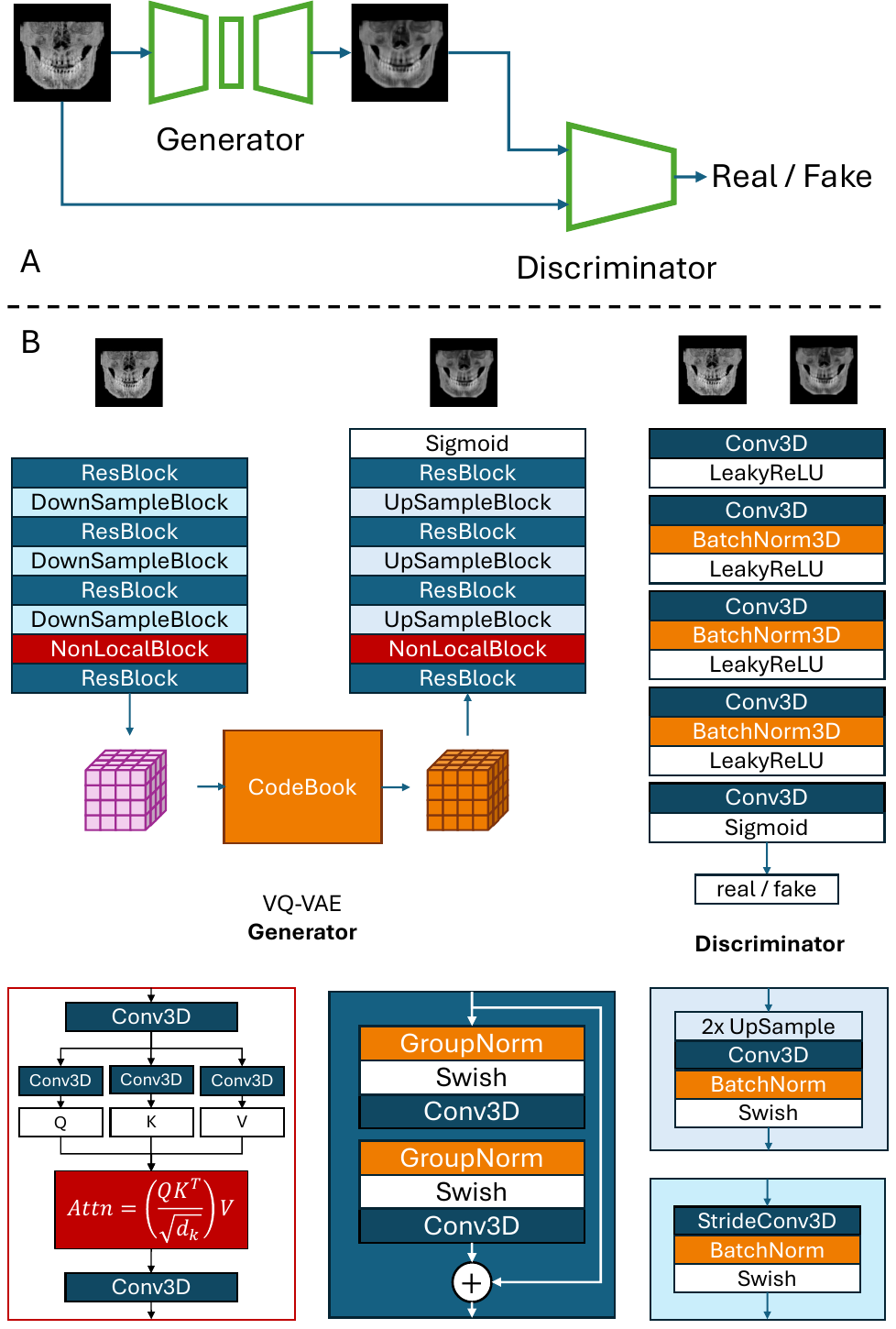}
    \caption{Detailed Network Structure Design of Masked VQ-GAN. A. The overall structure of a GAN, B. the detailed network structure and the layer definition.}
    \label{fig:gan}
\end{figure}

A VQ-GAN is used as the representation learning model in this task. The overall idea
is a generative adversarial networks, with the reconstruction model a Vector Quantized Variational Autoencoder (VQ-VAE).

\subsection{Network Structure}
The detailed structure is shown in \ref{fig:gan}B. The original VQ-GAN is proposed for 2D images, the network has been adapted to this 3D setting, while large followed the original structure, with some slight changes to avoid overfitting.

The encoder consists of residual blocks and downsample blocks, and at the lowest resolution, a non-local block is added to provide the interaction in global information. The decoder is designed in a reverse manner. The residual block is designed in a common manner, while the downsample and upsample block adds additional batch normalization and activation. Empirically, the network will not converge without the activation layer.

The discriminator has a similar structure to the PatchGAN discriminator, where the network does not output a single value indicating the real or fake of the whole image, but directly a patch of \(n \times n\).

\subsection{Loss Function}
The loss function is a combined loss between traditional GAN loss and the loss from VQ-VAE. In the paper proposed GAN, the penalty is simply binary cross-entropy, which is adopted in this work. Typically GAN loss can be written as:
\begin{align}
    \mathcal{L}_{G} &= -\mathbb{E}[\log D(G(x))] \\
    \mathcal{L}_{D} &= \mathbb{E}[\log D(x)] + \mathbb{E}[\log D(G(x))] \\
\end{align}
In VQ-GAN, a control term \(\lambda\) is introduced to balance the impact of adversarial loss and reconstruction loss:
\begin{align}
    \lambda = \frac{\nabla_{G_L}[\mathcal{L}_{recon}]}{\nabla_{G_L}[\mathcal{L}_{adv}] + \delta} \times 0.8
\end{align}
where \(G_L\) is the last trainable layer of the generator, \(\delta\) is a small constant to increase numerical stability.

Overall, the loss can be written as:
\begin{align}
    \mathcal{L}_G &= \mathcal{L}_{recon} + \mathcal{L}_{vq} + \lambda \mathcal{L}_{adv} \\
    \mathcal{L}_{recon} &= \| G(x) - x \|^2_2 \\
    \mathcal{L}_{vq} &= \|sg[f_e(x)]-e\|_2^2 + \beta \|f_e(x)-sg[e]\|_2^2 \\
    \mathcal{L}_{adv} &=  -\mathbb{E}[\log D(G(x))] \\
    \mathcal{L}_{D} &= \mathbb{E}[\log D(x)] + \mathbb{E}[\log D(G(x))] \\
\end{align}

This loss is used for both stage 1 and stage 2 training.

\subsection{Training Data}
In this project, the VQ-GAN does not directly reconstruct the raw CBCT image, but a recompiled segmentation image. This brought mainly two benefits. First, the model can learn reduced information from the segmentation map, without caring the detail such as bone and muscle internal structure and the neck. Second, this can totally eliminate the scan parameter discrepancy between the datasets, ensuring a successful reconstruction in the test set.

First, the raw image is segmented by DentalSegmentor into 6 regions, marked from 0 to 5, the content is shown in Table \ref{tab:region}. Then the non-zero part is added by 5 and divided by 10 to normalize and increase the contrast. Empirically, the VQ-GAN can easily converge on the processed 1-channel data, while struggling on the multi-channel segmentation map.
\begin{align}
    x' = (DS(x) + 5) / 10
\end{align}

For the teeth mask in stage two of the training, the teeth and mandibular canal region is selected is dilated as the target mask to simulate the missing teeth and bone around the jaw in ONJ.
\begin{align}
M_{teeth} &= 
\begin{cases} 
1 & \text{if } x' > 0.75 \\
0 & \text{otherwise} 
\end{cases}\\
M_{teeth} &= \text{MaxPool3D}(M_{teeth}, \text{kernel\_size}=5, \text{stride}=1, \text{padding}=2)\\
x'_{train} &= x' \odot (1 - M_{teeth})
\end{align}

\begin{table}
    \centering
    \begin{tabular}{|c|c|c|c|c|c|c|} \hline 
         Segmentaion&  0&  1&  2&  3&  4& 5\\ \hline 
         Class&  other&  upper skull&  mandible&  upper teeth&  lower teeth& mandibular canal\\ \hline
    \end{tabular}
    \caption{The segmented area generated by DentalSegmentor}
    \label{tab:region}
\end{table}

\subsection{Anomaly Subject Classification and Segmentation}
The model can not only be used to generate an anomaly map but also serve as a classifier to detect if an anomaly is presented in the image. The idea is similar but there are some detailed differences between these two methods.

For anomaly classification, the difference map is generated by calculating the absolute error \(\|G(x') - x'\|_1\). Then a threshold is applied to remove low difference, as low difference can be considered as the lack of expression ability of the model. After the thresholding, a binary erosion is used to remove single voxel difference, indicating small reconstruction error. Finally, the error is added up to form an anomaly score.

For anomaly segmentation, as the target is to generate as much as more anomaly area while minimizing the non-anomaly area, thresholding is not used, only erosion is used, because the edge of the anomaly area might contain differences with a low score. Additionally, dilation is used after erosion to recover the area eroded in the true anomaly region. 

\section{Experiments}
\label{sec:experiment}
\subsection{Dataset and Implementation Details}
The dataset used in this project is called "3D multimodal dental dataset based on CBCT and oral scan" posted in https://zenodo.org/records/10829675. It contained 290 scans from healthy subjects scanned with a spacing of \(0.25 \times 0.25 \times 0.25 \ mm^3\), where 287 of them were selected with the same dimensionality. Then the image is downsampled to a spacing of \(2 \times 2 \times 2 \ mm^3\), resulting \(75 \times 75 \times 52 \) in voxel image. The VQ-GAN in used reconstruct a image of \(64 \times 64 \times 64 \) and the original image is first padded to \(75 \times 75 \times 64 \) and then randomly cropped to the target size. During the final inference, a sliding window is implemented and the reconstruction is averaged through all the predictions. The dataset is split in 217 for training and 70 for inference. For the final model used for the application, the whole dataset is used for training.

Currently, only one patient's scan is available. It is scanned in a voxel size of \(0.4 \times 0.4 \times 0.4 \ mm^3\). The image gets the same preprocessing pipeline as the dataset.

The generator and discriminator are all optimized using Adam optimizer with a learning rate of 3e-4, using a batch size of 16. The training for the first stage is around 5000 epochs and second stage often coverage around 1000 epochs.

\subsection{Performance Analysis}
Two experiments are designed to validate the performance of the training method and model design. 

The first one is comparing the reconstruction performance of stages one and two with a baseline VQ-VAE, the comparison is conducted both quantitatively and qualitatively.

For the second experiment, I compare the anomaly detection performance between taking the difference between two reconstructions, and the difference between the original image and reconstruction. Two questions are to be answered: 1. Can the abnormal subject be successfully distinguished? 2. Is comparing to the original and reconstructed different?

\section{Results and discussion}
\label{sec:results}
\begin{figure}
    \centering
    \includegraphics[width=1\linewidth]{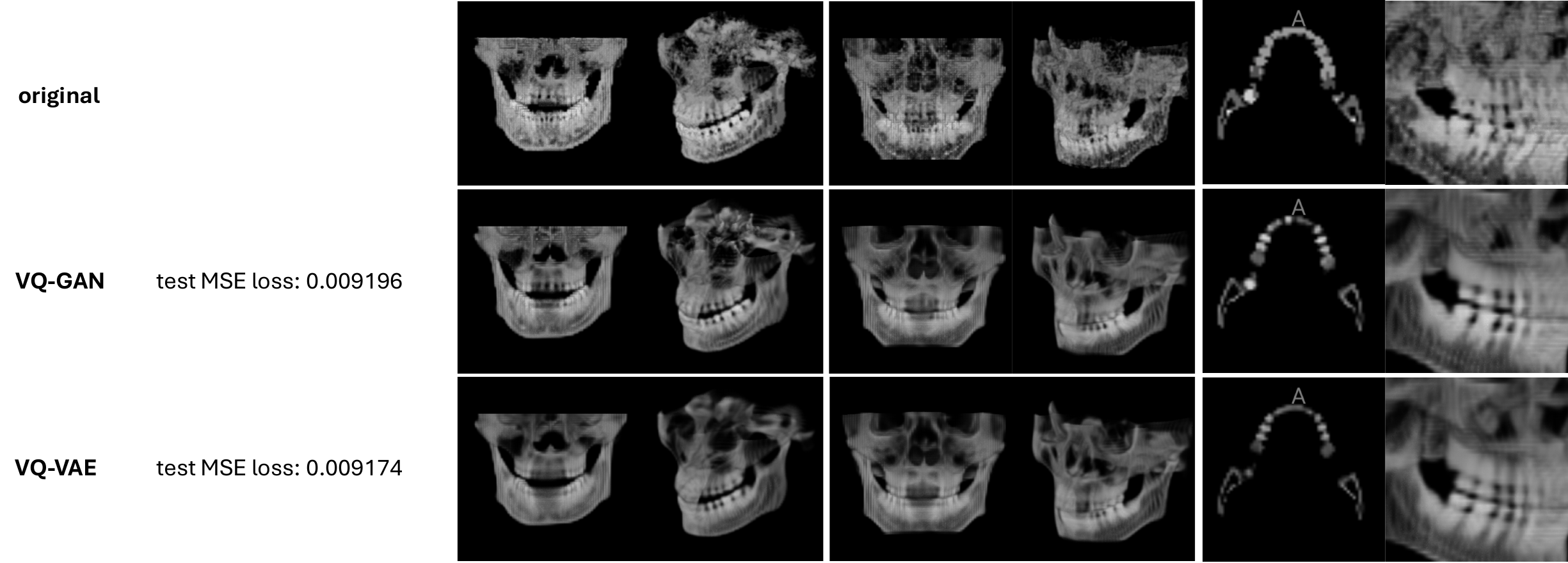}
    \caption{Reconstruction results in stage one. The VQ-GAN and VQ-VAE achieved similar test performance in MSE. While VQ-GAN can achieve slightly better fidelity in details such as teeth.}
    \label{fig:res_s1}
\end{figure}

\begin{figure}
    \centering
    \includegraphics[width=0.65\linewidth]{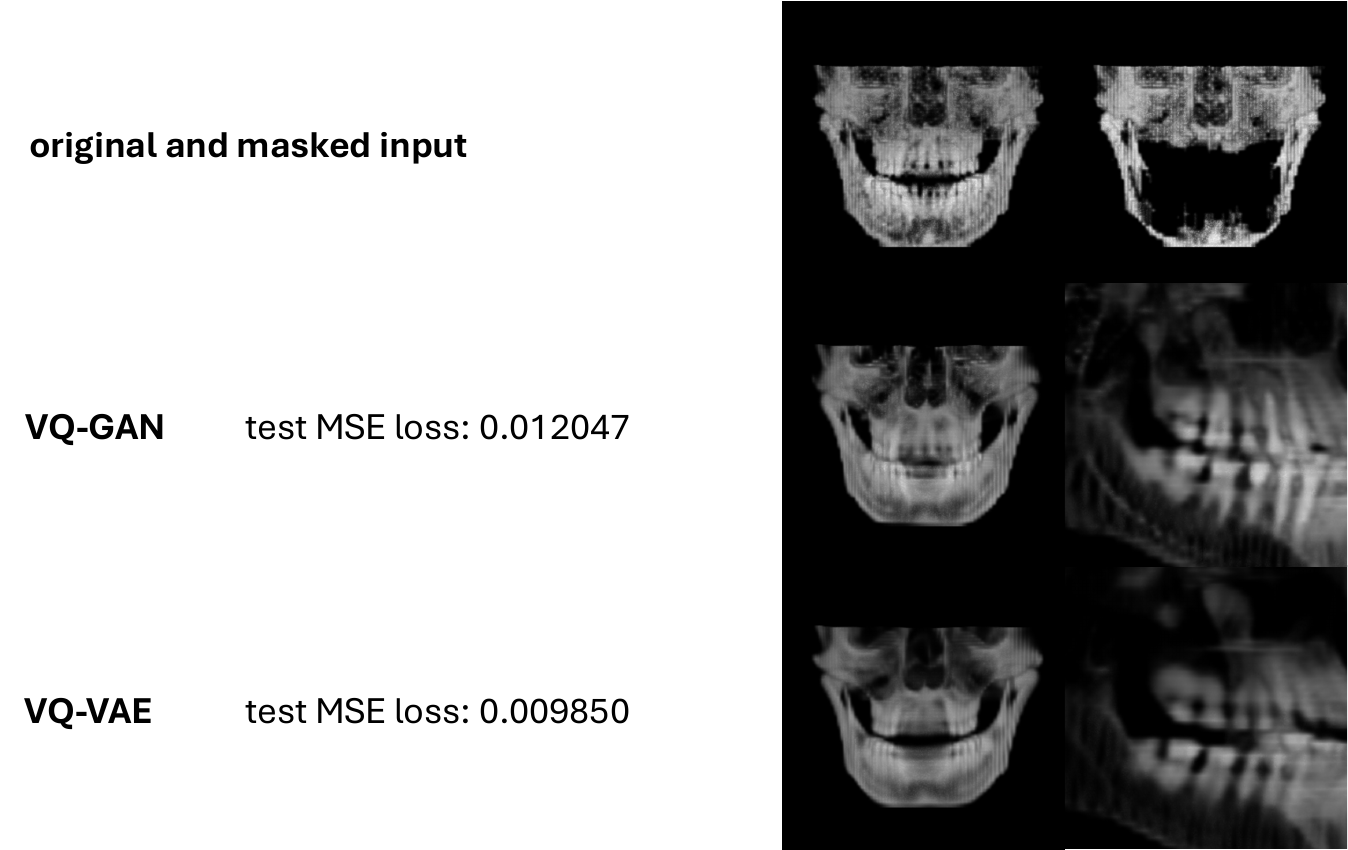}
    \caption{Reconstruction results in stage two. The VQ-VAE has better MSE. However, the prediction is clearly more blurred than the generation of VQ-GAN.}
    \label{fig:res_s2}
\end{figure}

\begin{figure}
    \centering
    \includegraphics[width=1\linewidth]{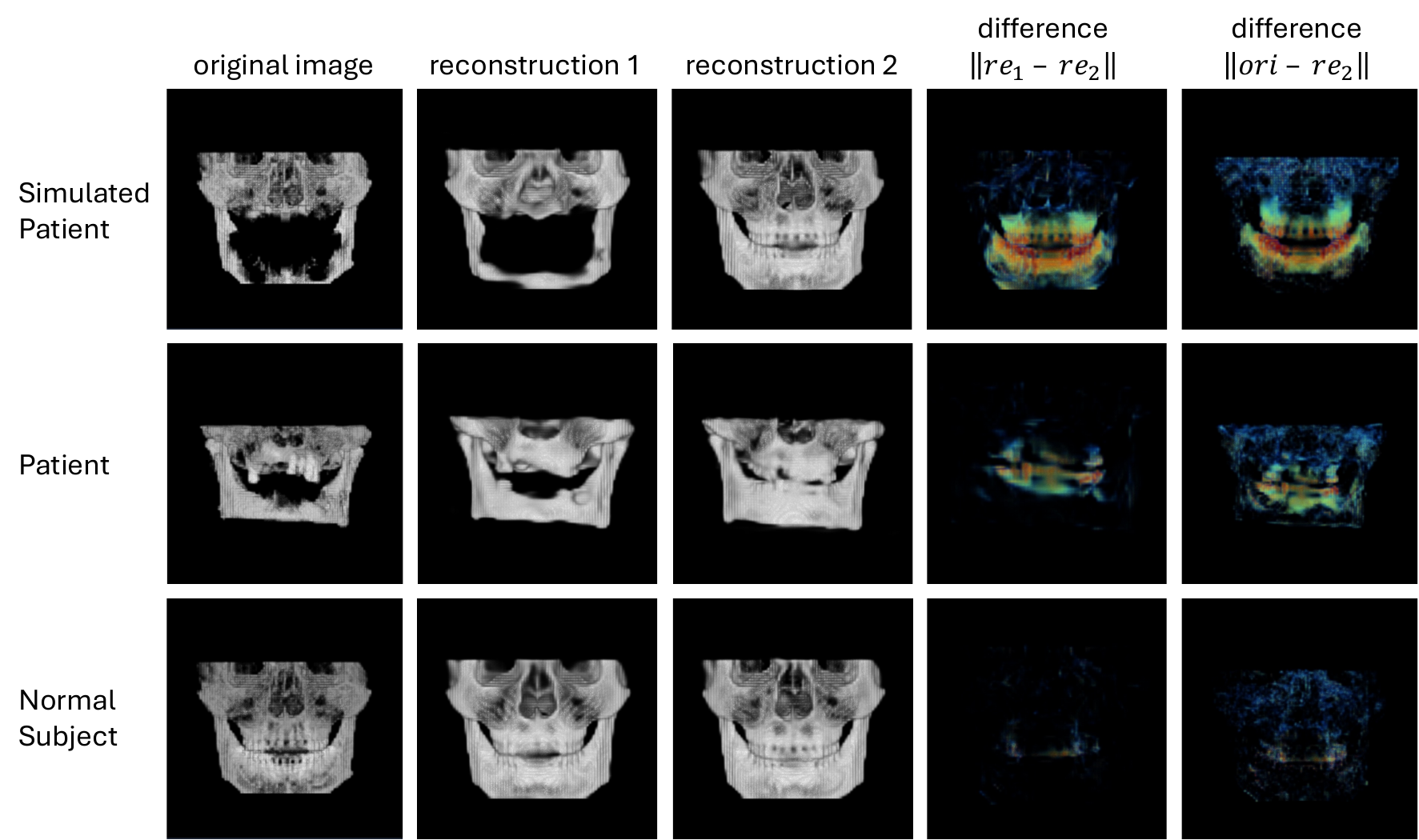}
    \caption{The anomaly reconstruction by two models trained on different stages, and the anomaly map generated by calculating the difference between reconstruction 2 and the original image, and between reconstruction 2 and 1.}
    \label{fig:ad_01}
\end{figure}

\subsection{Reconstruction performance}
The result of the stage one experiment is shown in Figure \ref{fig:res_s1}. The VQ-GAN has a reconstruction mean squared error of 0.009196 while VQ-VAE has an error of 0.009174. Both models can reconstruct the input image with high fidelity, with VQ-VAE obtaining a slightly better result. This might be due to the discrepancy in the training objectives between these two models. The loss function of VQ-VAE only contains the MSE loss and the vector quantization loss, while VQ-GAN should also try to fool the discriminator, optimizing the adversarial loss additionally. Nevertheless, this brings VQ-GAN a better fidelity than VQ-VAE. As is shown in the images, the reconstruction of the front teeth and teeth roots is clearer than the VQ-VAE.

The result of the stage two experiment is shown in Figure \ref{fig:res_s2}. The input image is a simulated patient with the teeth and jaw removed. The VQ-VAE shows a better MSE loss than VQ-GAN (0.009850, 0.012047 respectively). However, the image is even more blurred. While VQ-GAN can achieve the same fidelity. This is due to the loss of information in the teeth detail, there is no possibility to successfully guess the correct shape of the teeth. As a result, VQ-VAE chooses to predict a more blurred image to minimize the MSE loss, reflecting the expectation of the teeth. On the other hand, VQ-GAN has to output the image with high fidelity to fool the discriminator.

Overall, both models can perform well in this task, indicating the robustness of the training method, with VQ-VAE predicting the expectation value of voxels in the masked area and VQ-GAN predicting a sample of possible distribution.

\subsection{Anomaly detection performance}
The anomaly detection results are shown in Figure \ref{fig:ad_01}. The anomaly scores are also calculated for the whole test set. 

For the anomaly detection based on two reconstructions, only three normal subjects yielded an anomaly score of 1, while the rest had a score of 0. The patient received an anomaly score of 30, whereas simulated anomalies generally scored above 1000, indicating a clear demarcation between normal and pathological data. As illustrated in Figure \ref{fig:ad_01}, both anomalies in the simulated and real patients were highlighted with high confidence, and residuals were minimal.

For the anomaly detection based on the original image and the second reconstruction, 15 normal subjects exhibited anomaly scores greater than 0, with a maximum anomaly score of 8. The patient, however, received a score of 88, providing a slightly reduced contrast compared to the previous method, though still effective. Figure \ref{fig:ad_01} demonstrates that reconstruction error significantly influences the anomaly detection outcomes.

In conclusion, both methods successfully distinguished between patients and normal data, with the anomaly detection approach based on two reconstructions yielding superior results in terms of both anomaly scores and spatial anomaly maps.

\subsection{Discussion}
The experimental results underscore the efficacy and robustness of the proposed methodology. The framework is inherently flexible, allowing variations in model architecture and anomaly detection strategy. Since the latent space remains fixed during the second stage of training, it is possible to conduct anomaly detection directly within this latent representation or through a hybrid approach. As observed in Figure \ref{fig:ad_01}, the reconstruction of the real patient was less accurate compared to that of the simulated patient, likely due to limited spatial coverage in the upper skull region during scanning. This reduced coverage results in less available information, while the presence of teeth and jaw structures may introduce misleading signals for the generative model. Addressing this limitation may require enhancing the simulation process with a more sophisticated and detailed design.

However, as the proposed method is an unsupervised anomaly detection method, the missing teeth on the upper jaw are not classified as ORN. As a result, this method can only provide an initial segmentation but still needs human supervision.

\begin{figure}[ht]
    \centering
    \includegraphics[width=1\linewidth]{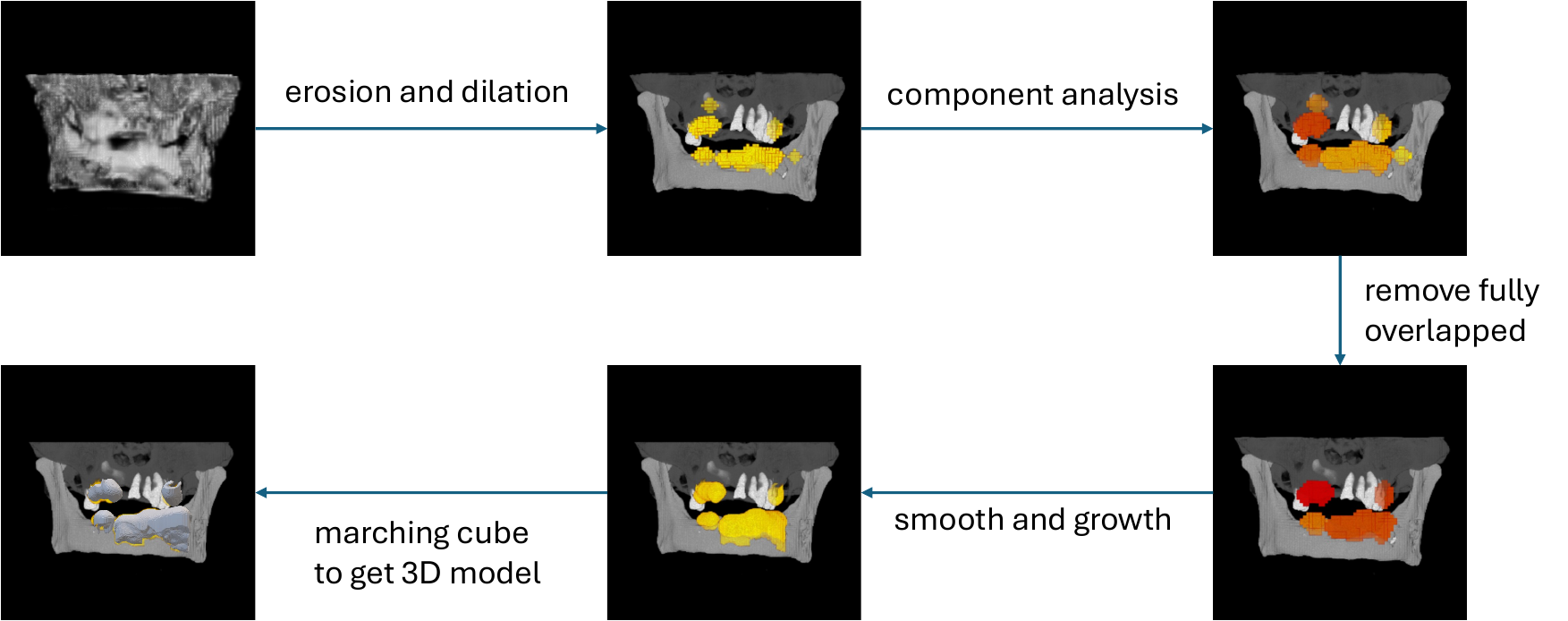}
    \caption{The post-processing pipeline. All codes are implemented using Python.}
    \label{fig:post-pro-piepline}
\end{figure}
\section{Post-processing}

\begin{figure}
    \centering
    \includegraphics[width=1\linewidth]{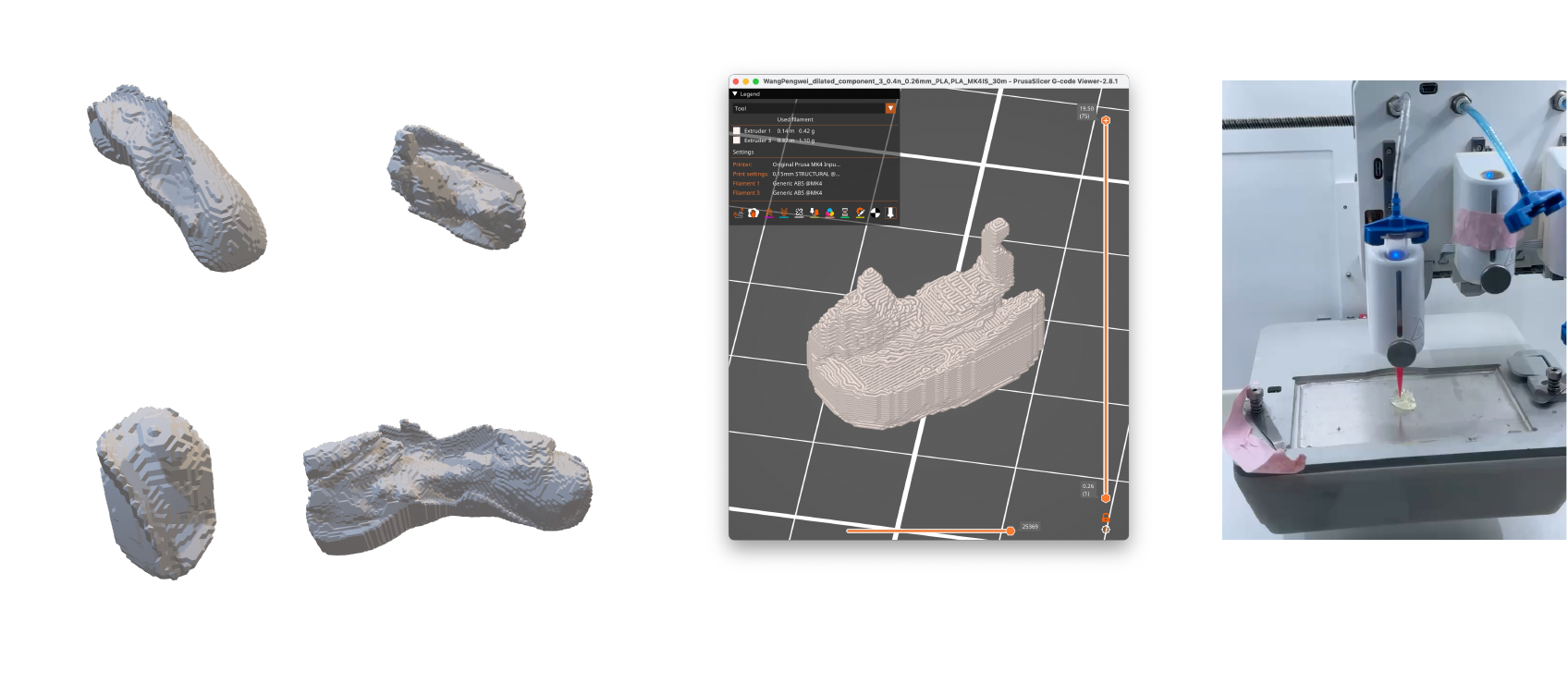}
    \caption{The generated STL model, 3d printer machine code, and the printer in printing.}
    \label{fig:post-pro-model}
\end{figure}

The overall post-processing pipeline is shown in Figure \ref{fig:post-pro-piepline} and Algorithm \ref{alg:post-processing}.

First, the reconstruction is generated and the raw difference map is calculated accordingly:
\begin{align}
    D = \| G_2(x') - G_1(x') \|_2^2
\end{align}
Then the map is processed with erosion and dilation to eliminate the small difference region and then applied with thresholding to get the target segmentation.

After getting the segmentation, component analysis is conducted to get independent possible wound regions. The component analysis is used to separate the disconnected parts in the segmentation. After the component analysis, all the disconnected region are stored separately for further processing.

Each region is checked if it has a major overlap with the segmentation. If true, it indicates that this will not be a wound area, it is removed accordingly. The region is then grown in a certain direction to fill up to empty area. Finally, the marching cube algorithm \cite{lorensen1998marching} is used to get an STL model from each region.

\begin{algorithm}
\caption{Post-Processing Pipeline}
\label{alg:post-processing}
\begin{algorithmic}[1]
    \State \textbf{Input:} Input data $x'$, Generator models $G_1$ and $G_2$
    \State \textbf{Output:} STL models of possible wound regions

    \State $D \gets \| G_2(x') - G_1(x') \|_2^2$

    \State $D \gets$ erosion($D$)
    \State $D \gets$ dilation($D$)
    \State $S \gets$ threshold($D$)

    \State $\{R_i\} \gets$ component\_analysis($S$)

    \For{$R_i$ in $\{R_i\}$}
        \If{overlap($R_i$, $S$)}
            \State remove($R_i$)
        \EndIf
        \State grow($R_i$)
    \EndFor

    \For{$R_i$ in $\{R_i\}$}
        \State STL\_model($R_i$) $\gets$ marching\_cubes($R_i$)
    \EndFor

    \State \Return STL models
\end{algorithmic}
\end{algorithm}

\begin{figure}[ht]
    \centering
    \includegraphics[width=0.9\linewidth]{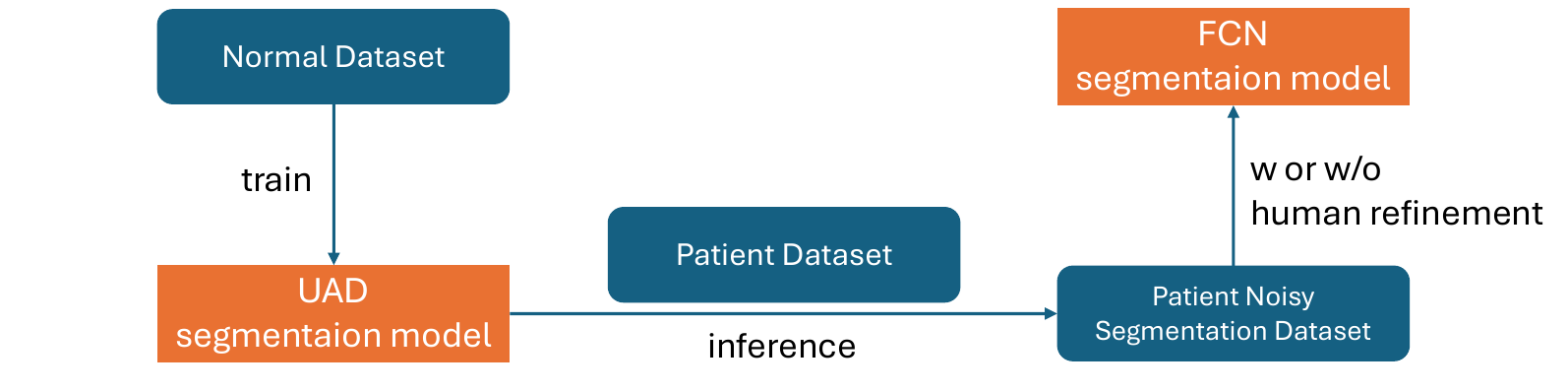}
    \caption{Potential pipeline for training a supervised segmentation model (Fully Connected Network, FCN) using noisy segmentation generated by UAD segmentation model.}
    \label{fig:future}
\end{figure}

\subsection{Avoid soft tissue}
\begin{figure}
    \centering
    \includegraphics[width=0.9\linewidth]{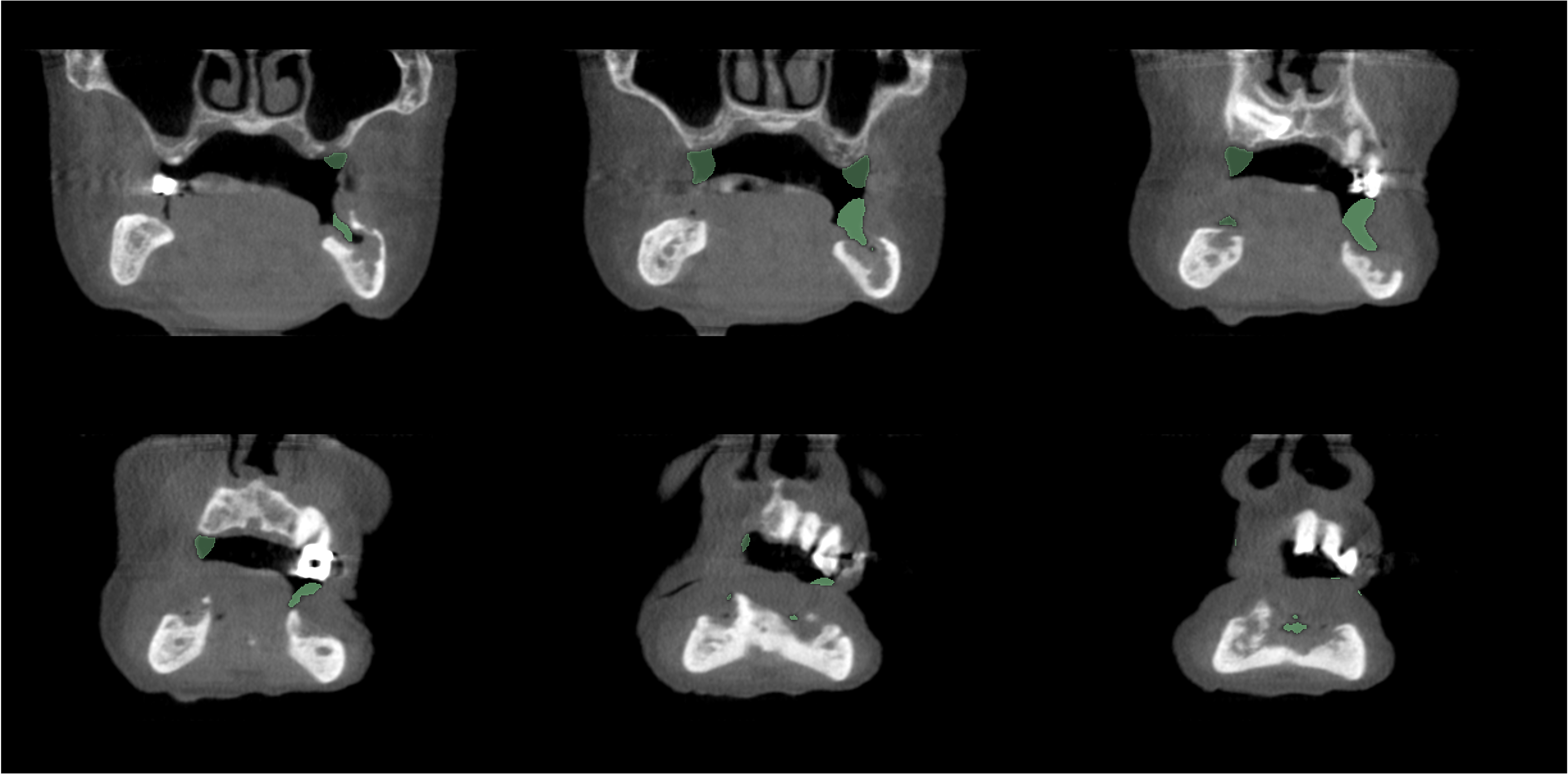}
    \caption{The segmentation area avoiding the soft tissue}
    \label{fig:avoid-soft}
\end{figure}

The previous analysis only considered the bone area and ignored the soft tissue. However, there is soft tissue covered on the broken bone. As a result, I also considered removing the overlapped area between the segmentation and soft tissue, and the result is shown in Figure \ref{fig:avoid-soft}.

I have noticed that there are holes inside the tissue. To finalize the anomaly area, expert suggestions are required.

\section{Conclusion and future work}
In this project, an unsupervised anomaly detection method is successfully implemented and used in osteoradionecrosis of the jaw to generate a 3D model of the wound. A novel two-stage training method is proposed and validated by experiments.

In the future, this model can be used to generate coarse segmentation for the training of a supervised fully connected network for more precise generation, as is shown in Figure \ref{fig:future}. Similar methods have been validated in white matter hyperintensity (WMH) lesions in brain MRI \cite{liu2024deepwmh}.

\begin{figure}
    \centering
    \includegraphics[width=1\linewidth]{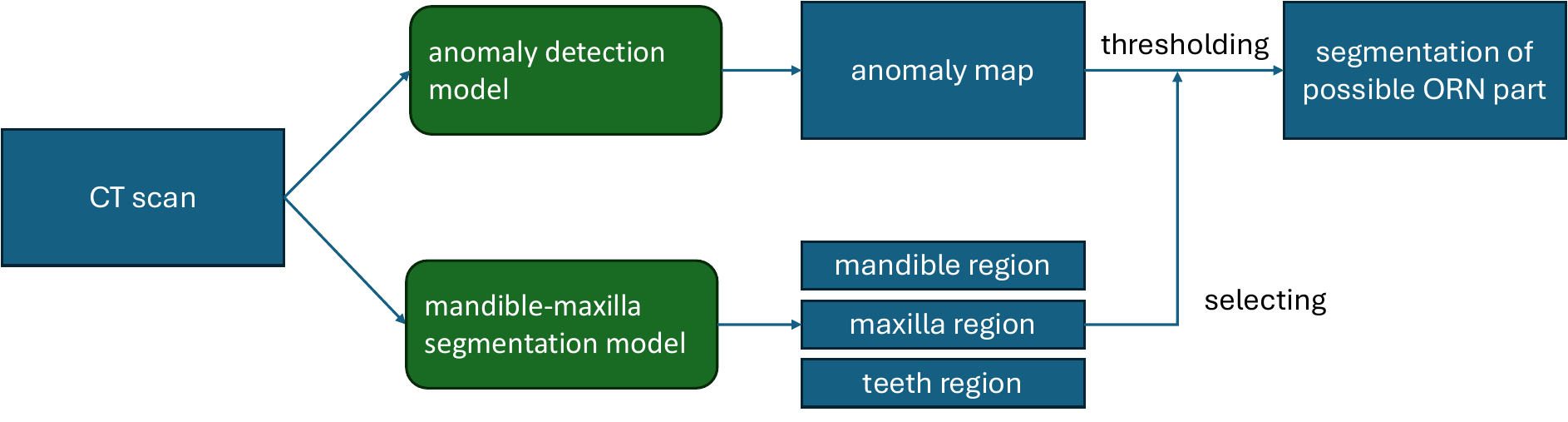}
    \caption{An alternative pipeline that can also possibly work.}
    \label{fig:alter-pipeline}
\end{figure}

\begin{figure}
    \centering
    \includegraphics[width=0.7\linewidth]{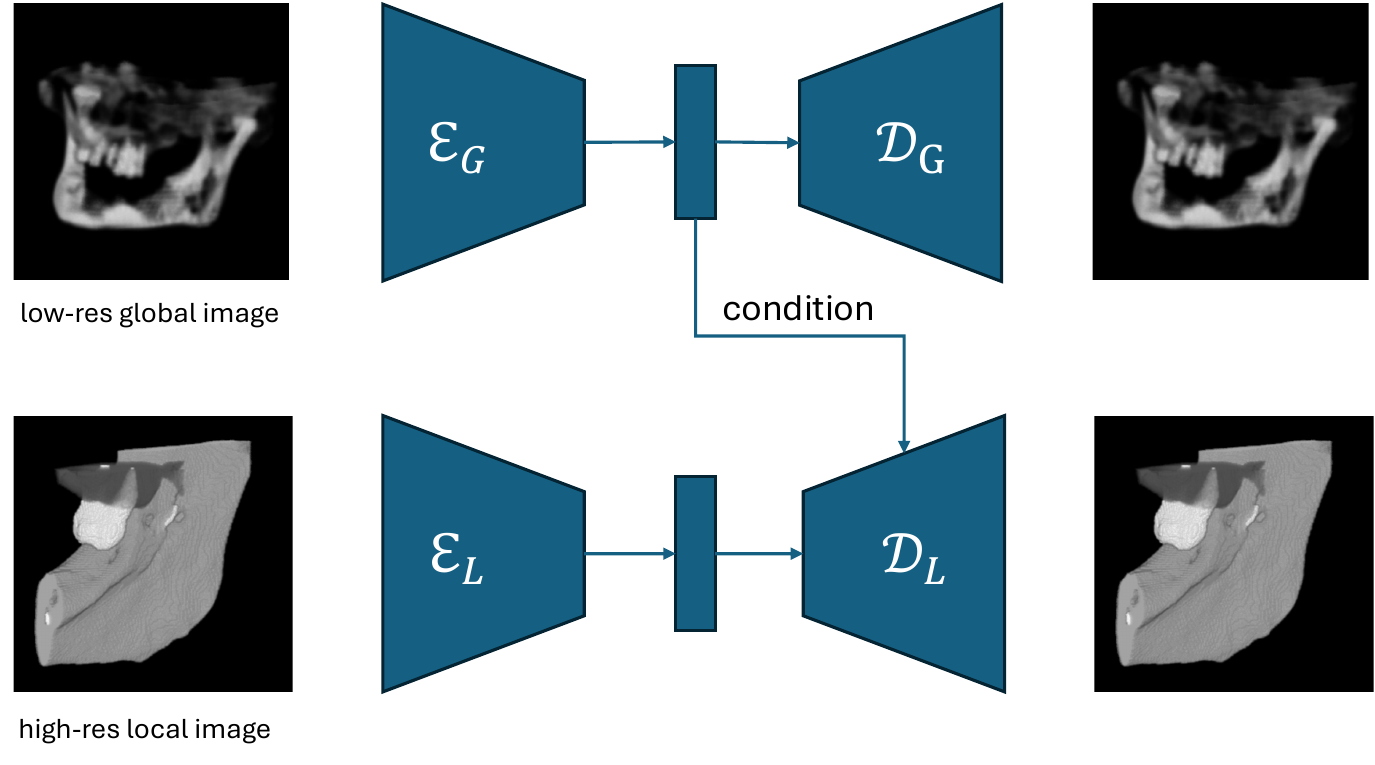}
    \caption{A possible network design to reconstruct the high resolution imaging with limited computational resources}
    \label{fig:glo-loc-net}
\end{figure}

The design choices of the preprocessing and network have not gone through rigorous ablation study. As a result, there might also be potential space for improvement in the network design. Potential improvements are: 1) directly reconstruct the original imaging instead of the segmentation map, as is shown in Figure \ref{fig:alter-pipeline}; 2) use a special network design to reconstruct the high-resolution version of the imaging, such as what is shown in Figure \ref{fig:glo-loc-net}.

Overall, to make a usable system, more work on the discussion with experts and system implementation should be considered. 

\section*{Acknowledgments}
I gratefully acknowledge the help from Kaiqi.

\bibliographystyle{unsrt}  
\bibliography{references}

\end{document}